\documentclass[twocolumn,times]{aastex62}

\usepackage{subfigure}
\usepackage{amsmath}

\newcommand{\Deg}{^{\circ}} %Needs $$

\submitjournal{\apjl}

\shorttitle{Energy Spect. Anisotropy of Cosmic Rays E$\geq$10$^{19.2}$~$eV$}
\shortauthors{Abbasi et al.}
\begin{document}

\title{Evidence of Intermediate-Scale Energy Spectrum Anisotropy of Cosmic Rays E$\geq$10$^{19.2}$~eV with the Telescope Array Surface Detector}

\correspondingauthor{J.P. Lundquist}
\email{jplundquist@cosmic.utah.edu}

\author[0000-0001-6141-4205]{R.U. Abbasi}
\affiliation{High Energy Astrophysics Institute and Department of Physics and Astronomy, University of Utah, Salt Lake City, Utah, USA}

\author{M. Abe}
\affiliation{The Graduate School of Science and Engineering, Saitama University, Saitama, Saitama, Japan}

\author{T. Abu-Zayyad}
\affiliation{High Energy Astrophysics Institute and Department of Physics and Astronomy, University of Utah, Salt Lake City, Utah, USA}

\author{M. Allen}
\affiliation{High Energy Astrophysics Institute and Department of Physics and Astronomy, University of Utah, Salt Lake City, Utah, USA}

\author{R. Azuma}
\affiliation{Graduate School of Science and Engineering, Tokyo Institute of Technology, Meguro, Tokyo, Japan}

\author{E. Barcikowski}
\affiliation{High Energy Astrophysics Institute and Department of Physics and Astronomy, University of Utah, Salt Lake City, Utah, USA}

\author{J.W. Belz}
\affiliation{High Energy Astrophysics Institute and Department of Physics and Astronomy, University of Utah, Salt Lake City, Utah, USA}

\author{D.R. Bergman}
\affiliation{High Energy Astrophysics Institute and Department of Physics and Astronomy, University of Utah, Salt Lake City, Utah, USA}

\author{S.A. Blake}
\affiliation{High Energy Astrophysics Institute and Department of Physics and Astronomy, University of Utah, Salt Lake City, Utah, USA}

\author{R. Cady}
\affiliation{High Energy Astrophysics Institute and Department of Physics and Astronomy, University of Utah, Salt Lake City, Utah, USA}

% ...

\begin{abstract}
An intermediate-scale energy spectrum anisotropy has been found in the arrival directions of ultra-high energy cosmic rays of energies above $10^{19.2}$~eV in the northern hemisphere, using 7 years of Telescope Array surface detector data. A relative energy distribution test is done comparing events inside oversampled spherical caps of equal exposure, to those outside, using the Poisson likelihood ratio. The center of maximum significance is at $9^h16^m$, 45$\Deg$, and has a deficit of events with energies $10^{19.2}$$\leq$E$<$$10^{19.75}$~eV and an excess for E$\geq$10$^{19.75}$~eV. The post-trial probability of this energy anisotropy, appearing by chance anywhere on an isotropic sky, is found by Monte Carlo simulation to be 9$\times$10$^{-5}$ (3.74$\sigma_{global}$).
\end{abstract}


%%%%%%%%%%%%%%%%%%%%%
\section{INTRODUCTION}
Though sources of ultra-high energy cosmic rays (UHECR) are still unknown, galactic origin is improbable due to the lack of strong anisotropy at energies above $10^{19}$~eV. Due to cosmic ray particle interactions with the infrared and microwave background radiation, the distribution of UHECR sources should be limited to distances smaller than 100 Mpc for protons and iron and 20 Mpc for intermediate mass nuclei like helium/oxygen/carbon/nitrogen (\cite{Kotera:2011cp}). The number of possible accelerators in this volume is limited by energy considerations to galaxy clusters, active galaxy jets and lobes, supermassive black holes (AGN’s), starburst galaxies, gamma-ray bursts, and magnetars.

These extragalactic objects are distributed along the local large scale structure (LSS), most evidently along the ``supergalactic plane.'' Nearby AGNs are concentrated around LSS with typical clustering lengths of 5--15 Mpc. The typical amplitude of such AGN concentrations is estimated to be a few hundred percent of the averaged density within a 20$\Deg$ radius (\cite{Ajello:2012yg}). This suggests that intermediate-scale anisotropy could have a similar angular scale.

Indeed, the Telescope Array (TA) experiment has observed evidence (at the 3.4$\sigma$ level) for a “Hotspot” near Ursa Major for event energies above 57~EeV (\cite{Abbasi:2014lda}). This anisotropy has a maximum significance in a 20 degree circle centered on $9^h48^m$, 43$\Deg$.

The present paper is an extension to lower energies (E$<$57~EeV) and is specifically a search for differences in the energy distribution of events within the field of view (FOV).

%%%%%%%%%%%%%%%%%%%%%
\section{EXPERIMENT}
The TA experiment in Millard County, Utah (39.3$\Deg$~N, 112.9$\Deg$ W) consists of a surface detector (SD) array (\cite{AbuZayyad:2012kk}) and three fluorescence detectors (FD) (\cite{Tokuno:2012mi}). The SD array has 507 plastic scintillation detectors, each 3 m$^2$ in area, placed on a 1.2 km spaced square grid resulting in a 700 km$^2$ collection area that makes it the largest cosmic-ray detector in the northern hemisphere. Data has been collected since 2008 with a close to 100\% duty cycle. Less than 10\% of SD data is observed in coincidence with the FD and is used to calibrate the SD energy scale using the calorimetric fluorescence technique.

%%%%%%%%%%%%%%%%%%%%%
\section{DATA SET}
For this analysis, SD data recorded between May 11 of 2008 to 2015 is used. Events are reconstructed in the same manner as the ``Hotspot'' analysis of \cite{Abbasi:2014lda} though the data set cuts are tighter to improve the zenith angle resolution at lower energies. The energy of reconstructed events is determined by SD and renormalized by 1/1.27 to match the calorimetrically determined energy scale of the FD (\cite{AbuZayyad:2012ru}).

Events are kept if they match the following criteria:
\begin{enumerate}
\item E$\geq$$10^{19.0}$~eV (where detection efficiency is $\sim$100\%).
\item Triggered at least four SDs.
\item Arrival direction zenith angle $<$55$\Deg$.
\item Reconstructed pointing direction error $<$5$\Deg$.
\item Core distance from array boundary $>$1.2 km.
\item Shower lateral distribution fit $\chi^2/dof$$<$10. 
\end{enumerate}

After cuts, there are a total of 3027 events in the data set.

The azimuthal angle distribution is in very good agreement with the theoretical flat distribution and the zenith angle distribution is in good agreement with the theoretical g($\theta$) = sin$(\theta)$cos$(\theta)$ distribution.The energy spectrum is also in good agreement with the published spectrum (\cite{AbuZayyad:2012ru}; \cite{TelescopeArray:2014zca}). 

The energy resolution and zenith angle resolution of events range from 10 to 20$\%$ and 1.0$\Deg$ to 1.5$\Deg$ respectively, depending on core distance from the array boundary and improve with increasing energy. These resolutions are sufficient to search for intermediate-scale anisotropy.

%%%%%%%%%%%%%%%%%%%%%
\section{MONTE CARLO DESCRIPTION} \label{sec:MC}
Each Monte Carlo (MC), and data, event is defined by their energy, zenith angle, azimuthal angle, and time. The latitude and longitude are defined from the center of TA at 39.3$\Deg$ Long., 112.9$\Deg$ Lat. Right Ascension (R.A.) and Declination (Dec.) equatorial coordinates are found using these variables (\cite{Vallado1991}).

Each MC set energy distribution is sampled by interpolation from a set of 386,125 MC events, with energies E$\geq$$10^{19.0}$~eV, reconstructed through an SD simulation that takes into account detector acceptance, on-time, and bias in the energy spectrum. This large MC set was created with the average HiRes spectrum (\cite{Abbasi:2007sv}) and was used for the TA spectrum measurement (\cite{Ivanov2012}). The same cuts applied to the data are applied to these simulated events.

The MC event sets have a uniform azimuth distribution and the geometrical zenith angle distribution of g($\theta$) = sin$(\theta)$cos$(\theta)$. On-time is simulated by randomly sampled trigger times from 246,499 data events with E$>$$10^{17.7}$~eV.

The result is that each set of isotropic MC events simulates the expected data given the detector configuration, and on-time, with no anisotropies. These MC sets are used to calculate the final significance of any data anisotropy.

%%%%%%%%%%%%%%%
\section{METHOD}
\subsection{Oversampling Anisotropy} \label{ssec:oversampling}
The oversampling method used in this paper is a modification of the large-scale anisotropy analysis developed by AGASA (\cite{Hayashida:1998qb}; \cite{Hayashida:1999ab}), namely an analysis done within overlapping spherical cap bins on the sky. The TA and HiRes collaborations have used similar methods previously (\cite{Abbasi:2014lda}; \cite{KawataICRC:2013}; \cite{2008ICRC....4..445I}). 

\subsubsection{Grid}
The oversampling is done on an equal opening angle grid with a median spacing of 0.5$\Deg$$\pm$$0.04$$\Deg$ between adjacent points. This spacing ensures equal sampling of the FOV and minimizes declination dependent bias. While the FOV extends to -16$\Deg$, the grid is stopped at 10$\Deg$ to avoid problems with the size of the spherical bins described in the next section.

\subsubsection{Equal Exposure Spherical Caps} \label{ssec:equalcaps}
There is a sample size bias in distribution tests of flux, such as $\chi^2$'s and likelihood ratios, which creates a declination bias in the calculated significances if the expectation sample size changes greatly with declination. Due to the zenith angle exposure g($\theta$) = sin$(\theta)$cos$(\theta)$ just such a bias is created if the spherical cap bin sizes are constant. An equal exposure binning is adopted such that the exposure ratio $\alpha = N_{on}/N_{off}$ (\cite{Gillessen:2004pm}) is a constant value at each grid point. 

A 2$\times$10$^7$ MC event set is used to determine the three parameter fit of the cap bin sizes, the average bin size (15.0$\Deg$, 20.0$\Deg$, 25.0$\Deg$, and 30.0$\Deg$), and the constant $\alpha$ exposure ratio that results in the required average bin size. After the bin sizes are found each exposure ratio $\alpha$ map is calculated from a 5$\times$10$^7$ MC event set to account for any remaining small variations from the bin size fit.

Smaller bin sizes do not have enough statistics inside them and larger bin sizes start to lose sufficient statistics outside for a distribution comparison. Also, a 35$\Deg$ bin size covers more than 50$\%$ of the oversampling grid and is no longer ``intermediate-scale.'' Furthermore, larger bin sizes have a greater change in shape at low declinations due to the exposure FOV cutoff. 

\begin{figure}[hb]
\centering
    \includegraphics[width=.497\textwidth]{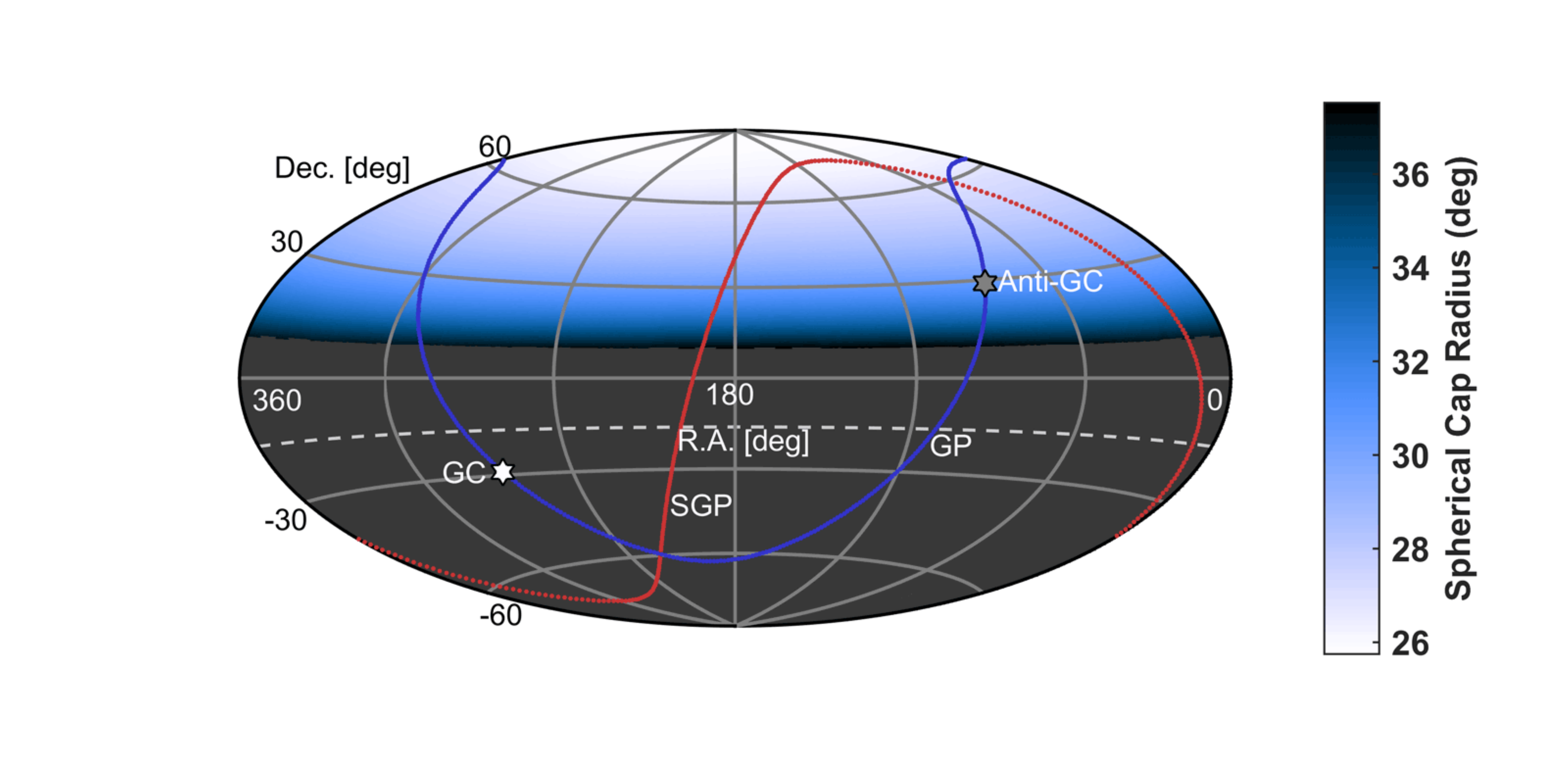}
  \caption{Equatorial Hammer-Aitoff projection of spherical cap bin sizes with an exposure ratio of $\alpha = 14.03\%$. The average bin radius is 30.0$\Deg$. The dashed curve at Dec. = -16$\Deg$ defines the FOV.}\label{fig:binsizes}
\end{figure}

Figure~\ref{fig:binsizes} shows the constant exposure ratio binning, $\alpha = 14.03\%$, that maximizes the data pre-trial significance which is an average bin size of 30.0$\Deg$. Ratios of $3.35\%$, $6.04\%$, $9.58\%$, and $14.03\%$ were tested to maximize the data pre-trial significance (the 15.0$\Deg$ to 30.0$\Deg$ spherical cap bin averages). This is a free parameter that the post-trial significance calculation takes into account. 

\subsection{Energy Distribution Comparison Test} \label{ssec:Etest}
The significance of a localized energy spectrum deviation is calculated using the binned Poisson likelihood ratio goodness of fit (GOF) test (\cite{Baker1984}; \cite{Agashe:2014kda}) to compare the energy distribution inside each spherical cap to the distribution of events outside the cap. This GOF test allows a low number of events in each energy bin, for both the observed ($N_{on}$ inside the bin), and expected ($N_{bg}$ normalized events outside) energy distributions.

Equation~\ref{eq:like} shows this test in terms of observed energy bin frequencies, $n_i$, expected frequencies, $\mu_i$, and exposure ratio $\alpha$. The local pre-trial $\sigma$ significance is calculated by approximating the likelihood ratio as $-\chi^2/2$ with degrees of freedom (DOF) $dof=\#bins$$+$$2$. The two additional DOF come from the estimated background and the combining of low statistic energy bins as described below. This was confirmed by MC simulation to follow the correct $\chi^2$ distribution.

\begin{subequations}
\begin{equation}
\chi^2 \simeq 2\sum_i \mu_i - n_i + n_i ln(n_i/\mu_i)
\label{eq:like}
\end{equation}
\begin{equation}
N_{on} = \sum_i n_i 
\end{equation}
\begin{equation}
N_{bg} = \sum_i \mu_i = \alpha(N_{events} - N_{on})
\label{eq:nbg}
\end{equation}
\end{subequations}

The choice of an energy bin width of 0.05~$\log_{10}$(E/eV) is $a\ priori$ based on the detector energy resolution and chosen to be slightly smaller than the average resolution for energies $10^{19}$$\leq$E$\leq$$10^{20.4}$~eV.

 The bias against the exact single bin $\chi^2$ distribution is less than +15$\%$ for $\mu_i$$>$2, and drops to +5$\%$ at expectations of 5 events in a bin (\cite{Heinrich2001}). If the expected number of events in an energy bin is less than 1 ($\mu_i$$<$1) it is combined with alternating adjacent bins. The resulting smallest energy bin expectations are greater than 2 ($\mu_i$$>$2). The combining of bins with $\mu_i$$<$1 ensures that the bias is positive for all bins instead of negative for the high energy bins with small expectations. This bias is smaller than other possible tests, is present for all locations on the sky map, and also present in the MC trials when calculating the global post-trial significance.

The expected energy spectrum is defined as the histogram of events outside the spherical cap bin ($N_{off}$) normalized to the expected number of events inside the cap bin ($N_{bg}$). The expected number of events inside the cap bins is calculated using the method of \cite{Li:1983fv}. 
   
At each point of the oversampling grid the exposure ratio $\alpha = N_{on}/N_{off}$ is calculated from a set of 5$\times$10$^7$ isotropic MC events. The background calculated from the data is $N_{bg}=$ $\alpha$$N_{off}$ $=$ $\alpha$$(N_{events}$$-$$N_{on})$ and therefore varies depending on the magnitude of $N_{on}$ inside each spherical cap bin (\cite{Gillessen:2004pm}).

The lowest energy threshold tested to maximize the pre-trial significance was $10^{19.0}$~eV as the detection efficiency is $\sim$100\% above this energy. Above $10^{19.4}$~eV there are only 546 events which is insufficient statistics. The maximum significance is for energies E$\geq$$10^{19.2}$~eV. This is a free parameter and appropriate penalty factors for this scan are taken, as described in Section~\ref{ssect:global}.

Above $10^{19.2}$~eV there are 1332 events in the data set; 1248 with energy $10^{19.2}$$\leq$E$<$$10^{19.75}$~eV and 84 with E$\geq$$10^{19.75}$~eV. The energy threshold of $10^{19.75}$~eV (more exactly 57~EeV) was used for the TA Hotspot analysis as determined by the AGN correlation results from the Pierre Auger Observatory (PAO) (\cite{Abu-Zayyad:2013vza}).

%%%%%%%%%%%%%%%
\section{RESULTS}
\subsection{Density Map}

\begin{figure}[h]
\centering
 \leavevmode
  \subfigure[]{%
    \includegraphics[width=.497\textwidth]{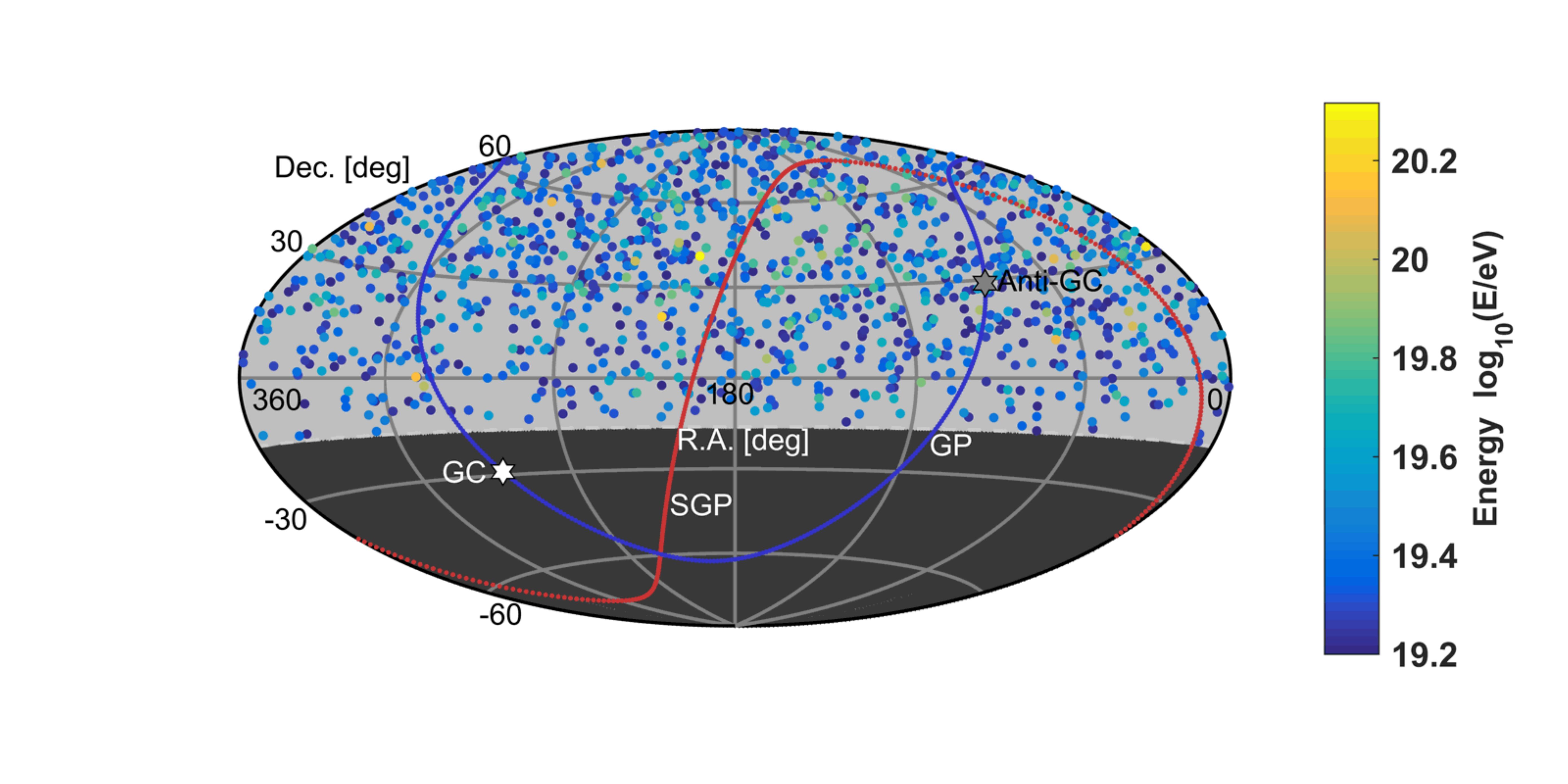}\label{fig:scatter}}
  \subfigure[]{%
    \includegraphics[width=.497\textwidth]{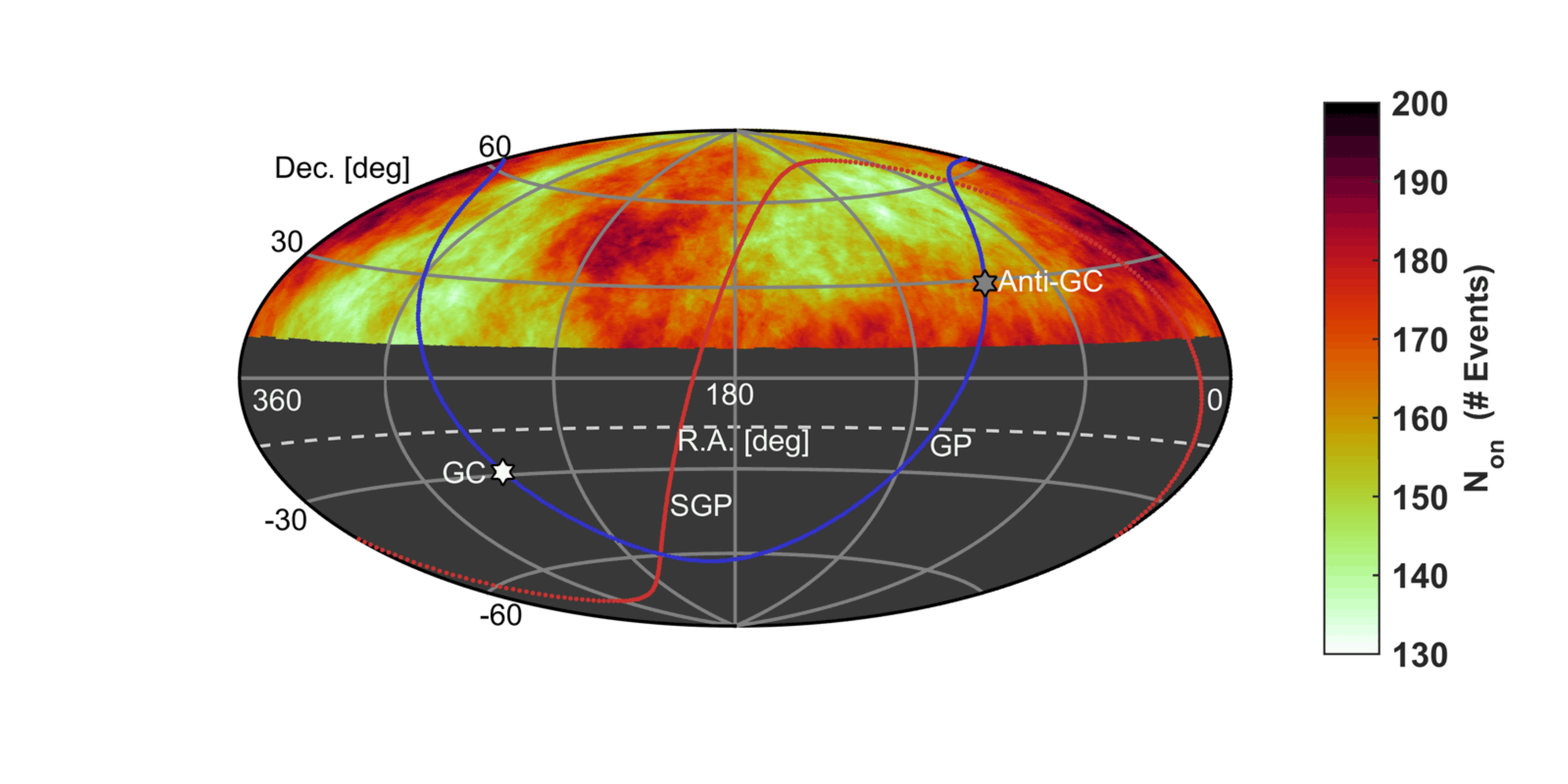}\label{fig:non}}
  \caption{Projections of UHECR events in the data set. (a) Scatter plot of events colored by $\log_{10}$(E/eV). (b) Number of observed events, $N_{on}$, at each grid point, inside 14.03\% equal exposure bins of the radius shown in Figure~\ref{fig:binsizes}. There is an event deficit at the previously reported Hotspot location ($9^h48^m$, 43$\Deg$). The dashed curve at Dec. = -16$\Deg$ defines the FOV. Solid curves indicate the galactic plane (GP) and supergalactic plane (SGP). White and grey hexagrams indicate the Galactic center (GC) and anti-galactic center (Anti-GC) respectively.}\label{fig:events}
\end{figure}

Figure~\ref{fig:scatter} shows an equatorial sky map of the 1332 cosmic-ray events observed by the TA SD with energies E$\geq$$10^{19.2}$~eV. Figure~\ref{fig:non} shows the oversampled number of events, $N_{on}$, using the equal opening angle sampling grid, and spherical cap bin size average of 30$\Deg$ as discussed in Section~\ref{ssec:oversampling}.

\subsection{Local Energy Anisotropy Significance} \label{ssec:local}
The pre-trial significance of local relative energy distribution deviations is calculated using the method of Section~\ref{ssec:Etest}. Inside each spherical cap bin the energy distribution of events ($N_{on}$) is compared to that outside ($N_{off}$) by the Poisson likelihood GOF test (Equation~\ref{eq:like}). The $\mu_i$ are the $N_{off}$ energy histogram frequencies normalized to the expected number of events ($N_{bg}$) by Equation~\ref{eq:nbg}. The $\alpha$ parameter is the exposure ratio described in Section~\ref{ssec:equalcaps}.

\begin{figure*}[b]
\centering
    \includegraphics[width=1\linewidth]{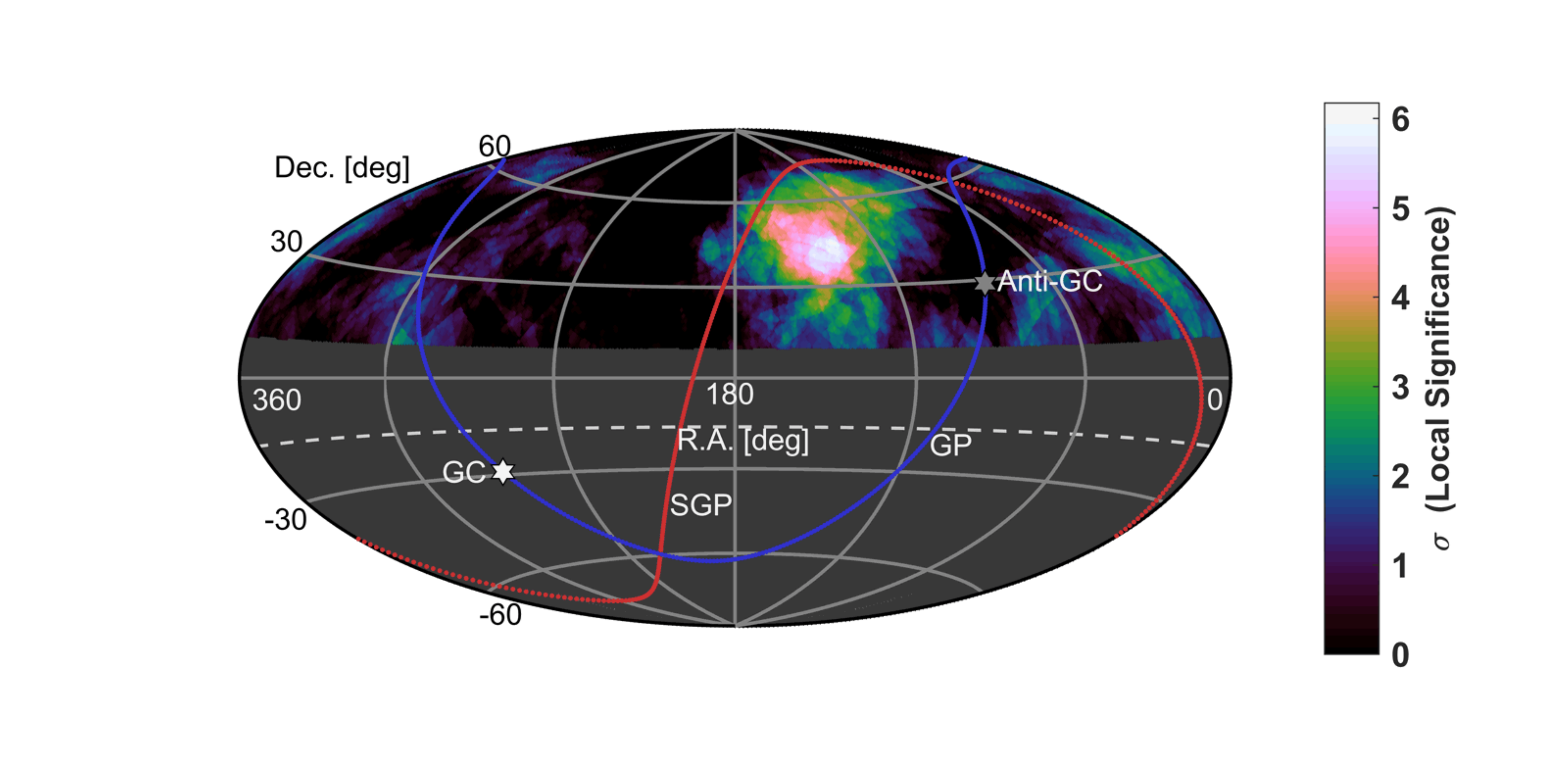}
  \caption{Projection of the local pre-trial energy spectrum anisotropy significance, for 14.03\% equal exposure spherical cap bins (E$\geq$10$^{19.2}$~eV). The maximum is 6.17$\sigma_{local}$ at $9^h16^m$, 45$\Deg$ and is 7$\Deg$ from the the Hotspot location of \cite{Abbasi:2014lda}. The dashed curve at Dec. = -16$\Deg$ defines the FOV. Solid curves indicate the galactic plane (GP) and supergalactic plane (SGP). White and grey hexagrams indicate the Galactic center (GC) and anti-galactic center (Anti-GC).}\label{fig:sigma}
\end{figure*}

The resulting local pre-trial energy anisotropy significance is shown in Figure~\ref{fig:sigma} using the spherical cap bin average of 30$\Deg$ and E$\geq$$10^{19.2}$~eV. The maximum pre-trial significance is 6.17$\sigma_{local}$ at $9^h16^m$, 45$\Deg$ inside a spherical cap bin of 28.43$\Deg$. This is 7$\Deg$ away from the Hotspot location of \cite{Abbasi:2014lda}.

\begin{figure*}[b]
\centering
 \subfigure[]{%
    \includegraphics[width=.497\textwidth]{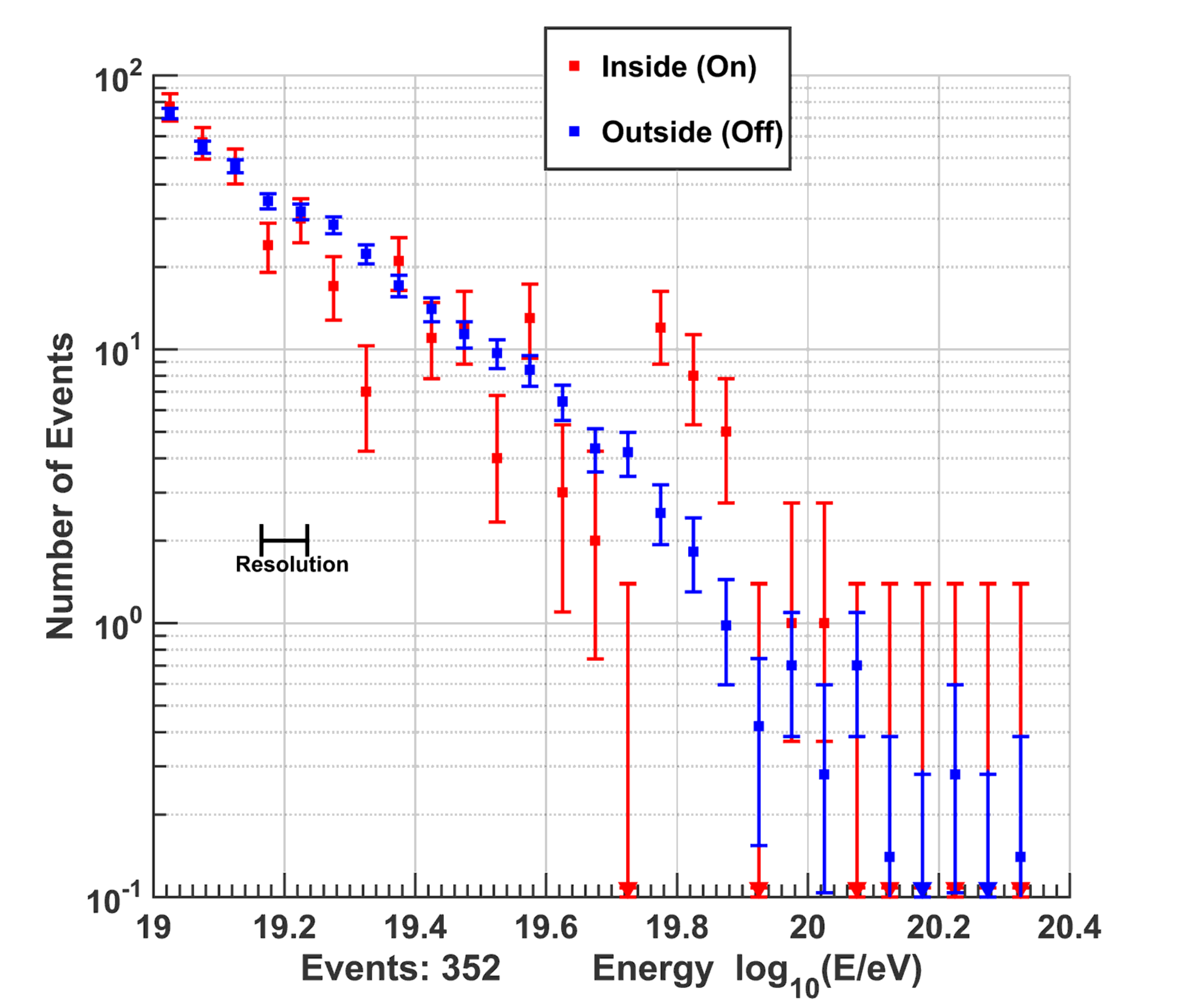}\label{fig:Ehistno}}
  \subfigure[]{%
    \includegraphics[width=.497\textwidth]{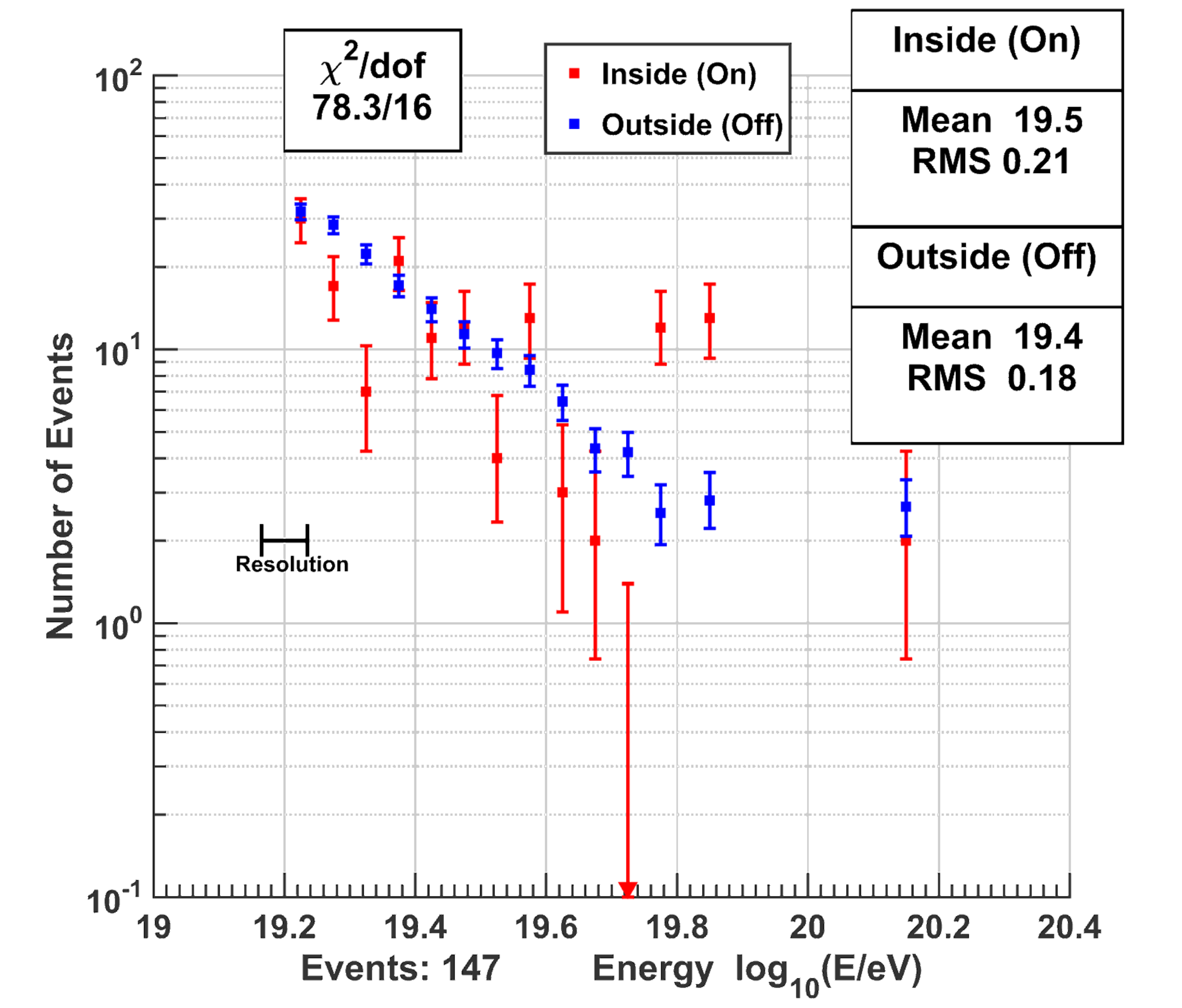}\label{fig:Ehist}}
  \caption{The maximum significance energy histograms of events inside the spherical cap bin of radius 28.43$\Deg$ (red) compared to the expected energies (blue) at $9^h16^m$, 45$\Deg$. (a) Before rebinning for events with energies E$>$10$^{19.0}$~eV. (b) After rebinning for energies E$>$10$^{19.2}$~eV (the maximum significance threshold). There are 147 events with an expectation of $N_{bg}$$=$$166.2$. Only three out of 11 bins for E$<$10$^{19.75}$~eV are above expectation.}\label{fig:ETestMaxSigma}
\end{figure*}

The histogram of events inside the spherical cap bin at maximum significance compared to the expected energies is shown in Figure~\ref{fig:ETestMaxSigma} with, and without, the rebinning discussed in Section~\ref{ssec:Etest}. Individual bin contributions to the statistical significance are from a Hotspot excess of events E$>$$10^{19.75}$~eV (27 observed, 8 expected, $\chi^2/dof = 38.1/4.5$), and a ``Coldspot'' deficit $10^{19.2}$$\leq$E$<$$10^{19.75}$~eV  (120 observed, 158 expected, $\chi^2/dof = 40.2/11.5$). The deficit is larger in magnitude than the excess as the expectation is $N_{bg}$$=$$166.2$ with an observed number of events $N_{on}$$=$$147$. 

\subsection{Global Significance} \label{ssect:global}
To calculate the global post-trial significance a scan penalty must be taken for the four energy thresholds (10$^{19.0}$, 10$^{19.1}$, 10$^{19.2}$, and 10$^{19.3}$~eV) and four exposure ratios (3.35$\%$, 6.04$\%$, 9.58$\%$, and 14.03$\%$) that were tested to maximize likelihood GOF $\sigma_{local}$ of Figure~\ref{fig:sigma}. 

Isotropic MC sets are made which have the same number of events as data for each energy threshold. The scanned variables are applied to each set to create 16 $\sigma_{local}$ maps. The maximum $\sigma_{local}$ significance at any grid point on all 16 maps is considered as one MC for counting MC sets that have a higher significance than the data. 

\begin{figure}[h]
\centering
    \includegraphics[width=.5\textwidth]{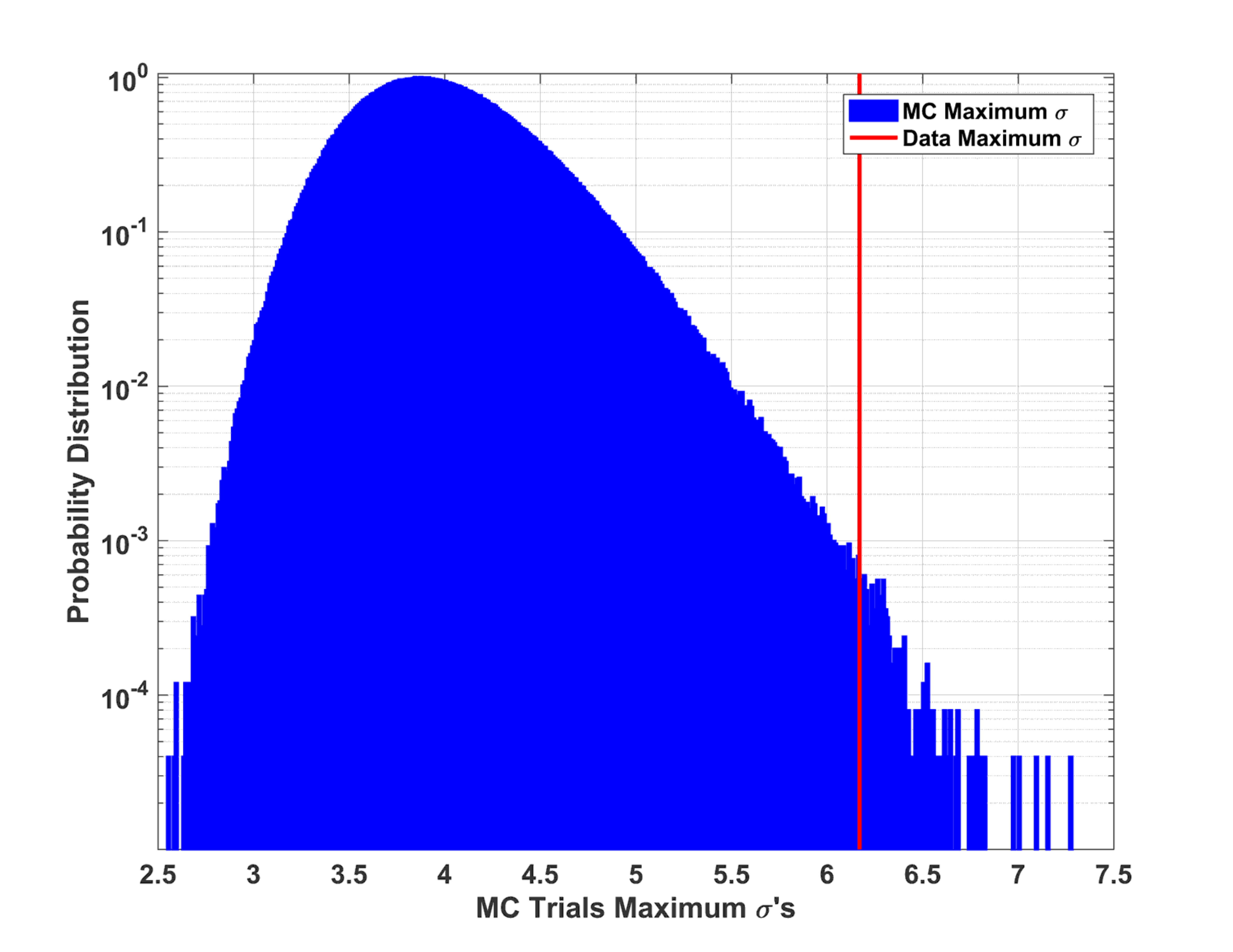}
  \caption{The distribution of the maximum local $\sigma$'s for 2.5$\times$10$^6$ MC trials. The area under the distribution above 6.17$\sigma_{local}$ corresponds to a 3.74$\sigma_{global}$ post-trial energy spectrum anisotropy significance.}\label{fig:sigpdf}
\end{figure}

Figure~\ref{fig:sigpdf} shows the distribution of the maximum $\sigma$'s of 2.5$\times$10$^6$ MC sets that are used to calculate the post-trial significance. There were 232 sets with a significance greater than 6.17$\sigma$. This corresponds to a global post-trial one-sided significance of 3.74$\sigma_{global}$.

\

\subsection{Systematic Checks} \label{ssec:sys}
There is a systematic bias on the energy determination due to seasonal and daily temperature induced changes to the average lateral distribution of particles in UHECR extensive air showers. This bias is estimated to fluctuate, about $\pm$7$\%$, with a negative bias in the winter months and positive in the summer. There's also an estimated fluctuation of about $\pm$5$\%$ throughout each 24 hour period. Applying these estimated energy corrections to the data results in a lowering of the local significance by about 0.05$\sigma$.

In the calculations of the equal exposure binning, the exposure ratio, and the global significance, the trigger times of events with energies E$\geq$10$^{17.7}$~eV were sampled to create the MC. This is to model how the TA SD would see an isotropic sky. It is known however that the acceptance, and therefore the trigger time distribution, is dependent on energy. To test the effect of this method MC sets were also created using uniform event trigger times and the calculation redone. The result is an increase in the pre-trial, and post-trial, significance of 0.04$\sigma$.

In addition to the seasonal energy correction test the energy distribution of events was also considered in anti-sidereal coordinates. This is an artificial coordinate system which emphasizes seasonal effects. No evidence for an energy spectrum anisotropy is found in anti-sidereal coordinates as would be expected for an anisotropy.

Other systematic checks include comparing the shower geometry variable (azimuth, zenith, core position etc.) distributions inside the anisotropic area, to outside, which show no disagreements (nor disagreements between different energy ranges inside the area) -- these distributions also agree with isotropic MC. The R.A., trigger time, and Dec. distributions inside the spherical cap are in good agreement between the Hotspot and Coldspot energy ranges -- they also each agree with isotropic MC. Also, the full energy distributions inside, and outside, the spherical cap do not show any significant seasonal variation.

%%%%%%%%%%%%%%%
\section{DISCUSSION}

The maximum energy anisotropy location is near the supergalactic plane that contains local galaxy clusters such as Ursa Major (20 Mpc from Earth), Coma (90 Mpc), and Virgo (20 Mpc). The closest distance between the hot/cold center and the supergalactic plane is 22$\Deg$ in the vicinity of Ursa Major and is 3$\Deg$ further than the Hotspot analysis. The difference is not statistically significant given the bin sizes and Gaussian fit to the Hotspot events as shown in \cite{Abbasi:2014lda}.

To get an idea if the measured energy spectrum anisotropy is correlated with the supergalatic plane the locations in Figure~\ref{fig:sigma} with excess/deficit behavior are converted to supergalactic coordinates and fit to a straight line (weighted by the pre-trial $\sigma^2$). The result corresponds to a great-circle rotated in declination by -16.5$\pm$0.1$\Deg$ tilted 2$\pm$1$\Deg$ around the center of the fit. This is suggestive of an extended feature that could be correlated with supergalactic structure. Possible mechanisms for producing such a shift include focusing of cosmic ray flux, for events with E$>$50~EeV, by supergalactic magnetic sheets as discussed in \cite{Biermann:1996xi}, and deflection of lower energy events transverse to the sheet as discussed in \cite{Ryu:1998up}. 

This feature may also be associated with the closest galaxy groups and/or the galaxy filament connected to the Virgo cluster (\cite{Dolag:2004fs}; \cite{He:2014mqa}; \cite{Pfeffer:2015idq}), if UHECR are protons as indicated by previous TA studies (\cite{Abbasi:2014sfa}). If the anisotropic UHECR are heavy nuclei, they may originate near the supergalactic plane and be deflected by extragalactic magnetic fields (EGMF) and the galactic magnetic halo field (GMF) (\cite{Tinyakov:2001ir}; \cite{Takami:2012uw}). If magnetic deflection or focusing is the mechanism, the magnitude is expected to be energy dependent.

To determine the origin of this feature, we will need greater UHECR statistics in the northern sky. Better information about the mass composition of the UHECRs, GMF, and EGMF are also important. The TA detector is currently being expanded by a factor of four (TAx4 \cite{SagawaICRC:2013}) and five years of additional data with this expanded detector should allow us to answer these questions.

%%%%%%%%%%%%%%%%%%%%%%%%%%%%
\section{SUMMARY}
Using seven years of TA SD ultra-high energy cosmic ray events a feature has been found appearing as a deficit of lower energy events (10$^{19.2}$$\leq$E$<$10$^{19.75}$~eV) and an excess of high energy events (E$\geq$10$^{19.75}$~eV) in the same region of the sky. The maximum local pre-trial significance is 6.17$\sigma$ and appears at $9^h16^m$, 45$\Deg$. The global post-trial probability of an energy spectrum anisotropy of this significance appearing by chance in an isotropic cosmic ray sky was found to be 9$\times$10$^{-5}$~(3.74$\sigma_{global}$). This feature is suggestive of energy dependent magnetic deflection of UHECR events.

%%%%%%%%%%%%%%%%%%%%%%%%%%%%%%%%%%%%%%%%%%%%%%%%%%%%%%%%%%%%%%%%%%%%%%%%%%%%%%%

\acknowledgments
The Telescope Array is supported by the Japan Society for
the Promotion of Science through Grants-in-Aid for Scientific Research
on Specially Promoted Research (21000002) ``Extreme Phenomena in the
Universe Explored by Highest Energy Cosmic Rays'' and for Scientific
Research (19104006), and the Inter-University Research Program of the
Institute for Cosmic Ray Research; by the U.S. National Science
Foundation awards PHY-0601915,
PHY-1404495, PHY-1404502, and PHY-1607727; 
by the National Research Foundation of Korea
(2015R1A2A1A01006870, 2015R1A2A1A15055344, 2016R1A5A1013277, 2007-0093860, 2016R1A2B4014967); by the Russian Academy of
Sciences, RFBR grant 16-02-00962a (INR), IISN project No. 4.4502.13,
and Belgian Science Policy under IUAP VII/37 (ULB). The foundations of
Dr. Ezekiel R. and Edna Wattis Dumke, Willard L. Eccles, and George
S. and Dolores Dor\'e Eccles all helped with generous donations. The
State of Utah provided support through its Economic Development
Board, and the University of Utah's Office of the Vice
President for Research. The experimental site became available through
the cooperation of the Utah School and Institutional Trust Lands
Administration (SITLA), U.S. Bureau of Land Management (BLM), and the
U.S. Air Force. We appreciate the assistance of the State of Utah and
Fillmore offices of the BLM in crafting the Plan of Development for
the site.  Patrick Shea assisted the collaboration with valuable advice. The people and officials of Millard County, 
Utah have provided greatly appreciated steadfast support. 
We are indebted to the Millard County Road Department for maintenance of the sites roads. 
We are grateful for the the contributions of our home institutions technical staffs and the University of Utah Center for High Performance Computing for the allocation of computer time.

\bibliographystyle{aasjournal}
%\bibliography{references}

\end{document}